\documentstyle[12pt,aps,psfig]{revtex}
\begin{document}
\title{ Transverse quasilinear relaxation in inhomogeneous
magnetic field}
\author{ Maxim Lyutikov}
\address{Theoretical Astrophysics, California Institute of Technology,
Pasadena, California 91125}
\date{25 January 1998}
\maketitle

\begin{abstract}
Transverse quasilinear relaxation of the cyclotron-Cherenkov
instability in the inhomogeneous magnetic field of pulsar
magnetospheres is considered. 
We find quasilinear states in which the kinetic 
cyclotron-Cherenkov instability of a beam propagating through strongly
magnetized pair plasma is saturated by the  force arising in the
inhomogeneous field due to the conservation of the adiabatic invariant.
The resulting wave intensities generally have nonpower law frequency
dependence, but in a broad frequency range can be well approximated by the 
power law with the spectral index $-2$.
The emergent spectra and fluxes are
consistent with the one observed from pulsars.
\end{abstract}

\section{Introduction}

In this paper we consider 
quasilinear relaxation in inhomogeneous
magnetic field of a highly relativistic beam propagating along the 
strong magnetic field through a pair plasma. 
This describes the physical conditions on the open field lines
of pulsar magnetospheres (e.g. \cite{Arons83}).
The possibility of the cyclotron-Cherenkov instability
of the beam 
in the pulsar magnetosphere has been suggested by 
\cite{KawamuraSuzuki} and \cite{LominadzeMachabeliMkhailovsky}
and developed later by \cite{MachabeliUsov1989},
 \cite{Kazbegi}, \cite{LyutikovPhD}.
The 
cyclotron-Cherenkov instability  develops at the anomalous
Doppler resonance
\begin{equation}
\, \omega({\bf k}) - k_{\parallel} v_{\parallel} -s
 {\, \omega_B \over  \gamma }
=0 \, \, \mbox{ for $s< 0$}
\label{ba}
\end{equation}
where $\, \omega({\bf k})$ is the frequency of the normal mode,
${\bf k}$ is a wave vector, $v$ is the velocity of the
resonant particle, $\, \omega_B= |q| B /mc $ is the nonrelativistic
gyrofrequency, $\,  \gamma$ is the Lorentz factor  in the
pulsar frame, $q$ is the charge of the resonant particle, $m$ is its
mass and $c$ is the speed of light.
It has been shown (e.g. \cite{LyutikovPhD}), that cyclotron-Cherenkov instability can 
explain a broad variety of the observed pulsar phenomena.

Close to the stellar surface, where the initial beam is produced
and accelerated, the particles  quickly  reach their
ground gyrational state due to the sychrotron emission in a superstrong
magnetic field, so that  their distribution becomes virtually
one dimensional \cite{Arons83}. 
In the outer parts of magnetosphere it becomes possible to 
satisfy the anomalous
 Doppler resonance -  the cyclotron-Cherenkov instability
develops  bringing about the diffusion of particles in transverse
moments. The relevant saturation mechanism then determines the 
final spectrum (which can be later modified be the absorption processes).

The nonlinear saturation of the cyclotron-Cherenkov instability
due to the diffusion of the resonant particles has been previously
considered by several authors.  Kawamura \& Suzuki
\cite{KawamuraSuzuki} neglected  the possible stabilizing effects 
of the radiation reaction force due to the 
cyclotron emission at the 
normal Doppler resonance and the force arizing in the inhomogeneous
magnetic field due to the conservation of the adiabatic invariant. 
These forces result in a saturation of the quasilinear diffusion.

Lominadze et al.
\cite{LominadzeMachabeliMkhailovsky} were the first to notice
the importance of the radiation reaction force 
due to the emission at the normal Doppler resonance
on the saturation of the quasilinear diffusion.
Unfortunately,  Lominadze et al. \cite{LominadzeMachabeliMkhailovsky} 
 used an expression for the
cyclotron damping rates which is applicable only for the nonrelativistic
transverse motions, when $ \gamma \psi$ ($psi $ is the pitch angle)
is much less than unity. In the pulsar magnetospheres the 
development of the cyclotron-Cherenkov instability
results in a diffusion of particles in transverse moments, quickly 
increasing there transverse energy to relativistic values.

In a review paper  Lominadze et al.
\cite{Machabeli3}
took a correct account of the radiation reaction force 
due to the emission at the normal Doppler resonance
and 
pointed out  the importance  of the  the force arizing in the inhomogeneous
magnetic field due to the conservation of the adiabatic invariant
( ${\bf G} $ force Eq. (\ref{4})).
When considering the deceleration of the beam  Lominadze et al.
\cite{Machabeli3} has incorrectly neglected the 
radiation reaction force  due to the emission at the anomalous 
 Doppler resonance
in comparison with the 
radiation reaction force  due to the emission at the normal 
 Doppler resonance.

In this work we reconsider and extend the treatment of the 
quasilinear stage of the cyclotron-Cherenkov instability.
We have found a  state, in which
the particles are constantly slowing down their parallel motion,
mainly due to the component along magnetic field  of the
radiation reaction force of emission at the
 anomalous
 Doppler resonance. At the same time they 
 keep the pitch angle almost constant due to the balance
of the force $ {\bf G}$ and 
the component  perpendicular to the magnetic field  of the
radiation reaction force of emission at the
 anomalous
 Doppler resonance.
We calculate the distribution function and the wave intensities for
such quasilinear state.

In the process of the quasilinear diffusion the initial beam looses
a large fraction of its initial energy $ \approx 10 \%$, which
is enough to explain the typical luminosities of pulsars.
Though the quasisteady wave intensities are not strictly power laws
(see Eq. (\ref{224a})), they can well approximated by a power  law
with a spectral index $ F(\nu) \propto -2 $
($F(\nu)$ is the spectral flux density in Janskys) which is very
close to the observed mean spectral index of $-1.6$
\cite{lorimer}.
The predicted spectra also show a turn off at the low frequencies
$ \nu \leq 300 MHz$
and a flattering of spectrum at large frequencies
$ \nu \geq 1 GHz$ which may be related  to the  possible 
turn-up in the flux densities at mm-wavelengths \cite{KramerXilouris}.

\section{Quasilinear Diffusion}
\label{QuasilinearDiffusion}

A particle moving in a dielectric medium in magnetic field
with the velocity larger than the velocity of light in a
medium is emitting electromagnetic waves at the 
anomalous 
 Doppler resonance ( $s<0$  in Eq. (\ref{ba})) and at the normal 
 Doppler resonance ( $s>0$  in Eq. (\ref{ba})).
The radiation reaction due to the emission at the 
normal
 Doppler resonance slows the particle's motion along magnetic field
and decreases its transverse momentum. 
The radiation reaction due to the  emission at the 
anomalous
 Doppler resonance increases its transverse  momentum
and  also slows the particle's motion along magnetic field
\cite{Ginzburg}.
As the particle propagates into the region of lower magnetic field, the
 force  $ {\bf G}$  decreases the  its transverse momentum
and increases the parallel momentum.
The stationary state in transverse moments may be reach when the
actions of the $ {\bf G}$ force and  
radiation reaction due to the  emission at the
normal
 Doppler resonance is balanced by the 
radiation reaction due to the  emission at the
anomalous
 Doppler resonance.
 
The quasisteady stage may also be considered in terms of a
detailed balance of for the particle transitions between the 
Landau levels. The quasisteady stage is reached when
the number of induced  transitions $up$ in Landau
 levels due to the emission at the anomalous
 Doppler resonance is balanced by the number of the spontaneous
transitions $down$ in Landau
 levels due to the emission at the anomalous
 Doppler resonance.

The equations describing the quasilinear diffusion in the magnetic field
are
\begin{eqnarray}
&&
{ d f({\bf p}) \over d t} = {1\over \sin \psi} 
{\partial \over \partial  \psi  } \left[ \sin \psi
\left( D_{\psi \psi} \,
{\partial \over \partial  \psi  } +
D_{\psi p} {\partial \over \partial  p  } \right)
f({\bf p}) \right]
\mbox{} \nonumber \\ \mbox{}
&&
\hskip .3 truein
{1\over  p^2   } {\partial \over \partial  p  }
\left[   p^2  \left( D_{  p \psi} 
{\partial \over \partial  \psi  } + 
D_{p p } {\partial \over \partial  p  } \right)
f({\bf p}) \right]
\mbox{} \label{102} \\ \mbox{}
&&
\left( \begin{array}{cc}
D_{\psi \psi} \\
D_{\psi p} = D_{ p \psi } \\
 D_{p p }  
 \end{array}   
\right) =\,
 \sum_{s<0} \int { d {\bf k} \over ( 2 \pi)^3 } 
 w(s, {\bf p},{\bf k}) n({\bf k})
\left(  \begin{array}{cc}
( \Delta  \psi )^2 \\
( \Delta  \psi ) ( \Delta p) \\
 ( \Delta p)^2
 \end{array} \right)
\mbox{} \label{103} \\ \mbox{}
&&
{ d n({\bf k})  \over d t} =
\sum_{s<0} \int d {\bf p}  w(s, {\bf p},{\bf k}) 
\left(  n({\bf k}) \hbar \left(
{\partial \over \partial  p} +
{\cos \psi - (k v /\omega) \cos \theta \over p \sin \psi } 
{\partial \over \partial \psi } \right)
f({\bf p})  \right)
\label{1}
\end{eqnarray}
Where
\begin{eqnarray}
&&
\Delta p = { \hbar \omega \over v}
\hskip .2 truein
\Delta \psi = { \hbar ( \omega \cos \psi - k_{\parallel} v) \over 
p  v \sin \psi }
\mbox{} \label{12} \\ \mbox{}
&&
n({\bf k}) = { E^2 ({\bf k}) \over  \hbar \omega({\bf k}) }
\mbox{} \label{13} \\ \mbox{}
&&
 w(s, {\bf p},{\bf k}) ={  8 \pi^2 q^2 R_E({\bf k}) \over
 \hbar \omega({\bf k}) } 
\left| {\bf e} ({\bf k}) \cdot {\bf V} (s, {\bf p},{\bf k})
\right|^2 \delta (   \omega({\bf k}) - s \omega_B/\gamma - k_{\parallel}
v_{\parallel} )
\mbox{} \label{14} \\ \mbox{}
&&
{\bf V} (s, {\bf p},{\bf k}) =
\left( v_{\perp} {s\over z} J_s(z) ,\,
-i \sigma s  v_{\perp} J_s(z)^{\prime}, v_{\parallel} J_s(z)\right)
\label{2}
\end{eqnarray}
where
$E ^2({\bf k})  d {\bf k} / ( 2 \pi)^3 $ is the energy density of the waves
in the unit element range of ${\bf k}$-space. 
In the Table \ref{Dimen1} we give the dimensions of the main used quantities.

In Eq. (\ref{102}) we neglected the spontaneous emission processes at the 
anomalous Doppler resonance and the induced emission processes at the 
normal Doppler resonance. The net effect of the 
spontaneous emission at the
normal Doppler resonance is treated  as a damping force 
acting on each
particle in  the Boltzman-type left hand side of  equation (\ref{3}).
To be exact, we should have treated the effects of 
spontaneous emission at the
normal Doppler resonance as  stochastic terms 
in the Fokker-Plank-type terms on the right hand side of 
equation (\ref{3}). But the fact that
 the emission at the normal Doppler resonance
occurs on very high 
frequencies at which the presence of a medium is unimportant
in the dispersion relation of the waves and can be neglected
allows one to integrate the corresponding terms over angles
and sum over harmonics to obtain a classical synchrotron 
radiation  damping force, that can be treated using 
the Boltzman approach.
Thus, 
the 
 total time derivative of the distribution function
is
\begin{equation}
{ d f({\bf p}) \over d t} =
{ \partial f({\bf p}) \over \partial  t} +
{\bf v} { \partial f({\bf p}) \over \partial  {\bf r}}+
{ \partial \over \partial {\bf p} }  \left[  \left(
{\bf G + F } + {q\over c} \left( {\bf v} \times  {\bf B_0} \right) \right)
 f({\bf p})
\right]
\label{3}
\end{equation}
where ${\bf G} $ is the force due to the conservation of the 
adiabatic invariant
\begin{equation}
 G _{\parallel} =- \beta  \gamma \psi^2,
\hskip .2 truein
G _  {\perp} = - \beta \gamma \psi,
\hskip .2 truein 
\beta = { m c^2 \over R_B}
\label{4}
\end{equation}
Here $ R_B \approx 10^9 $ cm 
 is the radius of curvature and 
${\bf F }  $ is the radiation damping force
 due to the spontaneous synchrotron emission at the
normal Doppler resonance:
\begin{equation}
F_{\parallel} =- \alpha \gamma^2 \psi^2 ,
\hskip .2 truein
F_{\perp} = - \alpha \psi \left( 1+  \gamma^2 \psi^2\right)  ,
\hskip .2 truein
\alpha = { 2 q^2 \omega_B^2 \over 3 c^2}
\label{41}
\end{equation}

From (\ref{4}) and (\ref{5}) we find that
\begin{equation}
{  F_{\parallel} \over G _{\parallel} } = {\alpha \over \beta}
\gamma, 
\hskip .2 truein
{  F _{\perp} \over G_ {\perp} } = {\alpha \over \beta}
\gamma \psi^2, \,\, \mbox{ for $ \gamma \psi \gg 1$}
\label{5}
\end{equation}
where $r_L =c/\omega_B$ is a Larmor radius and $r_e =q^2/(m c^2) = 
2.8 \times 10^{-13} $cm  is 
a classical radius of an electron.

The dimensionless ration in (\ref{5}) is
\begin{equation}
{\alpha \over \beta}= 
{ 2 R_B r_e  \over r_L ^2} = 5 \times 10^{-4} \, R_{B,9} R_9 ^{-6} 
\label{6}
\end{equation}
$R_{B,9}=  R_B/10^9 {\rm cm}$ 
is the radius of curvature in units of $10^9$ cm, 
$R_9 =  R/10^9 {\rm cm}$
is the distance from the neutron star surface in units of $10^9$ cm.

Using (\ref{6}) we find that for the primary particles with $\gamma 
\approx 10^7$ 
\begin{eqnarray}
&&
{  F_{\parallel} \over G _{\parallel} } \gg 1 
\mbox{} \nonumber \\ \mbox{}
&&
{  F _{\perp} \over G_ {\perp} } \ll 1,
\hskip .2 truein
\mbox{ for $ \psi \ll \sqrt{ {  r_L ^2 \over  2 R_B r_e \gamma }} 
\approx   10^{-2} $ }
\label{7}
\end{eqnarray}

Then the total derivative (\ref{3}) may be written as 
\begin{equation}
{ d f({\bf p}) \over d t} =
{ \partial f({\bf p}) \over \partial  t} +
{\bf v} { \partial f({\bf p}) \over  \partial {\bf r}}+
{ 1\over p \sin \psi} { \partial  \over \partial \psi} 
\left( \sin \psi G_ {\perp} f({\bf p}) \right) +{ 1\over p^2}
 { \partial  \over  \partial p}
\left( p^2 F_{\parallel} f({\bf p}) \right)
\label{7a}
\end{equation}

We are interested in the quasilinear diffusion of the particles due to the
 resonant interaction with the waves at the anomalous
Doppler effect. We expand the transition currents 
(\ref{2}) in small $v_{\perp}$ keeping only $s=-1$ terms:
$ {\bf V} (-1,{\bf p},{\bf k}) = v_{\perp}/2 \left( 1, i \sigma ,0 \right)$.
Then for the waves propagating along magnetic field $ {\bf e} ({\bf k}) =
(1,0,0) $ we find 
\begin{equation}
 w(\pm 1, {\bf p},{\bf k}) ={  \pi^2 q^2 v_{\perp} ^2  \over
 \hbar \omega({\bf k}) }
 \delta (   \omega({\bf k}) - s \omega_B/\gamma - k_{\parallel}
v_{\parallel} )
\label{8}
\end{equation}
where we took into account that $ R_E({\bf k}) \approx 1/2$.

We now can find the diffusion coefficients in the approximation of
a one dimensional spectrum of the waves.
\begin{equation}
  n({\bf k}) = { 2 \pi \delta( \theta)  \over   k^2  \sin \theta   }  n( k ),
\hskip .2 truein
 n( k ) = \int d{\bf \Omega_k} {k^2 \over ( 2 \pi)^2 n({\bf k})}
\label{9}
\end{equation}

We first note that we can simplify the change in the pitch angle (\ref{12}) 
in the limit $  \psi ^2 \ll \delta$ and $ 1/\gamma^2 \ll \delta$ 
\begin{equation}
\Delta \psi \approx -  { \hbar \omega  \delta \over p v \sin \psi } 
\label{10}
\end{equation}
We then find
\begin{equation}
\left(  \begin{array}{cc}
D_{\psi \psi}\\
D_{\psi p} = D_{ p \psi } \\
D_{p p } \end{array} \right) =\,
\left(  \begin{array}{cc}
 \phantom{ {{{ {a\over b} \over {a\over b} } } \over {{ {a\over b} \over
{a\over b} } } } }
D \mbox{{\large $  { \delta  \over \gamma^2} $}}  E_k^2
 \left. \right|_ { k =k_{res} } \\
 \phantom{ {{{ {a\over b} \over {a\over b} } } \over {{ {a\over b} \over
{a\over b} } } } } -
 D \mbox{{\large $
 {  \psi m c \over \gamma} $}}  E_k^2
\left. \right|_ { k =k_{res} } \\
   \phantom{ {{{ {a\over b} \over {a\over b} } } \over {{ {a\over b} \over
{a\over b} } } } }
D \mbox{{\large $ { \psi ^2 m^2 c^2 \over \delta } $}}  E_k^2
\left. \right|_ { k =k_{res} }
 \end{array} \right),
\hskip .2 truein
D = { \pi^2 q^2  \over  m ^2 c^3 }  =
{ \pi ^2  r_e \over    m c  } \hskip .1 truein
\label{11}
\end{equation}
where 
\begin{equation}
E_k^2  = \hbar \omega(k) n(k) = \int {k^2  d {\bf \Omega}\over ( 2 \pi)^2}
 \hbar \omega({\bf k}) 
n({\bf k}) 
\label{1001}
\end{equation}
is energy density per unit of one-dimensional wave vector
and we assumed that $\omega({\bf k})$ is an isotropic function of ${\bf k}$.

We next solve the partial differential equation describing
the evolution of the distribution function by sucsessive approximations.
We first expand equation (\ref{102}) in small $\psi$ assuming that 
$ \partial/\partial \psi \simeq 1/\psi$. 
We also neglect the convection term assuming that the characteristic
time for the development of the quasilinear diffusion is much smaller
that the dynamic time of the plasma flow.
Then we assume that it is possible to separate the distribution
function into the parts depending on the $\psi$ and $p$:
\begin{equation}
f({\bf p}) =  Y (\psi, p) f(p)
\label{12a}
\end{equation}
with 
\begin{equation}
 f(p) =  2  \pi \int \sin \psi d \psi f({\bf p}),
\hskip .2 truein
\int  d p p^2  f(p) =  1
\label{121}
\end{equation}

In the lowest order in $\psi$ we obtain an  equation:
\begin{equation}
-
{ 1\over p \sin \psi} { \partial  \over \partial \psi}
\left( \sin \psi G_{\perp} Y (\psi) \right) =
{1\over \sin \psi}
{\partial \over \partial  \psi  } \left[ \sin \psi
 D _{\psi \psi}  \,
{\partial  Y (\psi)  \over \partial  \psi  } \right]
\label{13a}
\end{equation}
which has a solution
\begin{equation}
 Y (\psi) = { 1\over \pi \psi_0^2} 
 \exp\left( - {  \psi  ^2 \over  \psi_0^2 } \right),
\hskip .2 truein
\psi_0^2 = { D m c \delta E_k^2 \over \beta \gamma^2}=
 { D R_B  \delta  E_k^2 \over c \gamma^2}=
{\pi^2 \delta  R_B r_e E_k^2 \over \gamma^2 m c^2}
\label{15}
\end{equation}

The next order in $\psi$ gives
\begin{equation}
{ \partial f({\bf p}) \over \partial  t} +
 { \partial  \over  \partial p}
\left( F_{\parallel} f({\bf p}) \right)
=  {1\over \sin \psi}
{\partial \over \partial  \psi  } \left[ \sin \psi
D_{\psi p} {\partial f({\bf p}) \over \partial  p  } \right]
+
{1\over  p^2   } {\partial \over \partial  p  }
\left[   p^2 D_{p \psi } {\partial f({\bf p}) \over \partial  \psi  } \right]
\label{16}
\end{equation}

By integrating (\ref{16}) over $\psi$  with a weight  $\psi$ we find
the equation for the parallel distribution function:
\begin{equation}
{ \partial f( p) \over \partial  t} 
-
{\partial \over \partial  p  } \left( {A} E_k^2
 \gamma^2 f( p)
\right)  =
 {2 \over p^2} {\partial \over \partial  p  }
\left( p { D} m^2 c^2  E_k^2 f( p) \right)
\label{17}
\end{equation}
where 
\begin{equation}
 {A} = {\alpha \psi_0^2 \over E_k^2} =
 { 2  q^2 \omega_B^2 \psi_0^2  \over  3 c^2 E_k^2}= 
{ 2  \pi^2 \omega_B^2 q^4 R_B \delta \over 3\gamma^2 m^2 c^6}= 
{  2 \pi^2 R_B r_e^2 \delta \over 3  \gamma ^2 r_L^2} 
\label{18}
\end{equation}

The term containing ${A}$ describes the slowing of the particles due to the
radiation reaction force and the term  containing 
$D$ describes the slowing of the particles due to the
quasilinear diffusion, or, equivalently, due to the 
radiation reaction force of the anomalous Doppler resonance.
To estimate the relative importance of these terms we consider a ratio
\begin{equation}
{ {A} \gamma^3  \over {D} m c } =  {  \alpha \delta  \gamma \over \beta } =
{ 2 \delta  \gamma \over 3    }  { R_B r_e \over r_L^2}  \ll 1  
\label{19}
\end{equation}

Neglecting the second term on the left hand side of (\ref{19}) we find
\begin{equation}
{ \partial f( p) \over \partial  t} 
 -
 {2 \over p^2} {\partial \over \partial  p  }
\left( p { D} m^2 c^2  E_k^2 f( p) \right)
==0
\label{221}
\end{equation}

If the cyclotron quasilinear diffusion has time to fully
develop and reach a steady state, then the distribution function
of the resonant particles is
\begin{equation}
f( p) \propto { 1\over  p \,E_k^2 }
\label{222}
\end{equation}

Next we turn to the equation describing the temporal evolution
of the wave intensity (\ref{1}).
Neglecting the spontaneous emission term  and the wave convection
we find
\begin{equation}
{ \partial E_k^2  \over \partial  t} = -  \Gamma
 E_k^2  f(\gamma)_{res}
\label{223}
\end{equation}
where 
\begin{equation}
\Gamma = {1 \over f(\gamma)_{res}}\,
\sum_s \int d {\bf p}  w(s, {\bf p},{\bf k})
\left( \hbar \left(
{\partial \over \partial  p} +
{\cos \psi - (k v /\omega) \cos \theta \over p \sin \psi }
{\partial \over \partial \psi } \right)
f({\bf p})  \right)
\label{223a}
\end{equation}
and we introduced 
\begin{equation}
f(\gamma) \gamma ^2  d \gamma = f(p) p^2 d p
\label{223a1}
\end{equation}

We will estimate this growth rate for the emission along the external
magnetic field for  distribution (\ref{121}), (\ref{15}).
Neglecting $\partial/ \partial  p$ and assuming that $\psi^2 
\ll 2 \delta$ (so that most of the particles are moving with the
superluminal velocity) we find for $s=-1$
\begin{equation}
\Gamma = {\pi \omega_{p,res}^2 \over 2 \omega}
  \gamma^2 _{res}
\label{223b}
\end{equation}
(It is important to note that in the limit 
$\psi^2
\ll 2 \delta$ the growth rate does not depend
on the scatter in pitch angles).

Equations (\ref{221})  and (\ref{223b}) may be combined to a quasilinear
expression
\begin{equation}
{ \partial \over \partial  t}
\left(  f( \gamma) +  {2 \over p^2} {\partial \over \partial  p  }
\left( { p { D} m^2 c^2  E_k^2 \over \Gamma } \right)  \right)
=0
\label{2232}
\end{equation}

Which after integration gives
\begin{equation}
 f( \gamma) -  {2 \over \gamma^2} {\partial \over \partial  \gamma  }
\left( { \gamma { D}  E_k^2 \over \Gamma } \right)
= f^0( \gamma)
\label{224}
\end{equation}
Neglecting the initial density of particles in the region of quasilinear
relaxation and using 
 Eqs. (\ref{222}) and (\ref{224}) we can find a distribution
function and the asymptotic spectral shape:
\begin{eqnarray}
&& f(\gamma) = { 1\over 2 \gamma^3 } 
\left( { 1\over \log(\gamma_{max}/\gamma)  \log (\gamma_{max}/\gamma_{min}) }
\right)^{1/2}
\mbox{} \label{2241} \\ \mbox{}
&&
 E_k^2 = { m c^2 \delta r_L \gamma^2 \over 2 \pi r_e r_S^2}
\left( { \log(\gamma_{max}/\gamma)  \over  \log (\gamma_{max}/\gamma_{min}) }
\right)^{1/2}      
=
 { m c^4 \delta^3 \over 2 \pi \omega^2 r_e r_L r_S^2}
\left( {  \log({ \gamma_{max} \omega  r_L /( c \delta ) }) 
\over  \log( \gamma_{max}  / \gamma_{min} )  }
\right)^{1/2}
\label{224a}
\end{eqnarray}

\begin{figure}
\psfig{file=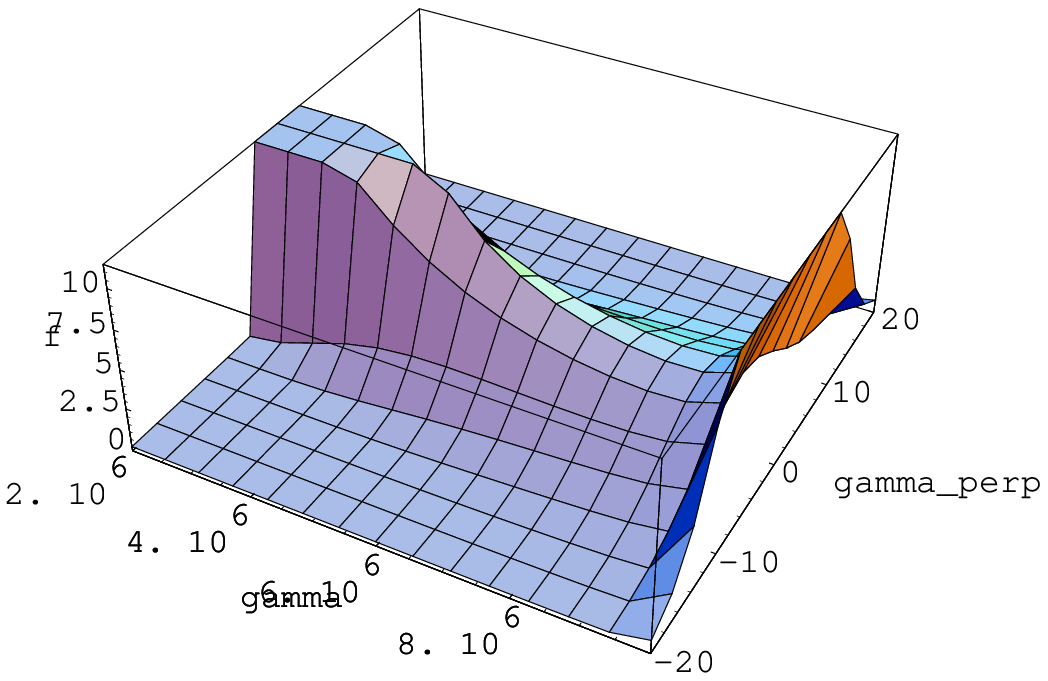,width=15.0cm}
\psfig{file=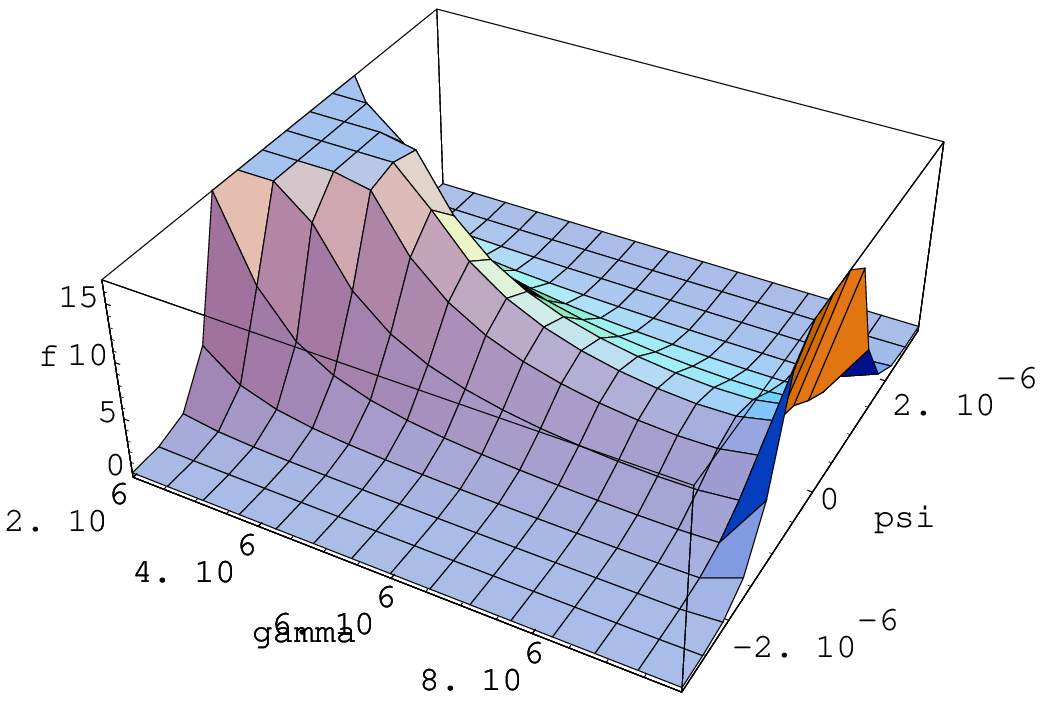,width=15.0cm}
\caption{
Asymptotic distribution functions in $\gamma - p_{\perp}$ and 
 $\gamma - \psi$ spaces in arbitrary units for $\gamma_{max} =10^7$. 
The spike at  the $\gamma=\gamma_{max}$
is an artifact of the initial distribution function $f(\gamma) ^0=
\delta(\gamma- \gamma_{max})$.
The divergence at $\gamma= \gamma_{max}$ is weak (logarithmic) 
and would be removed if the more realistic 
nitial distribution function was used.
\label{Distr}}
\end{figure}

\begin{figure}
\psfig{file=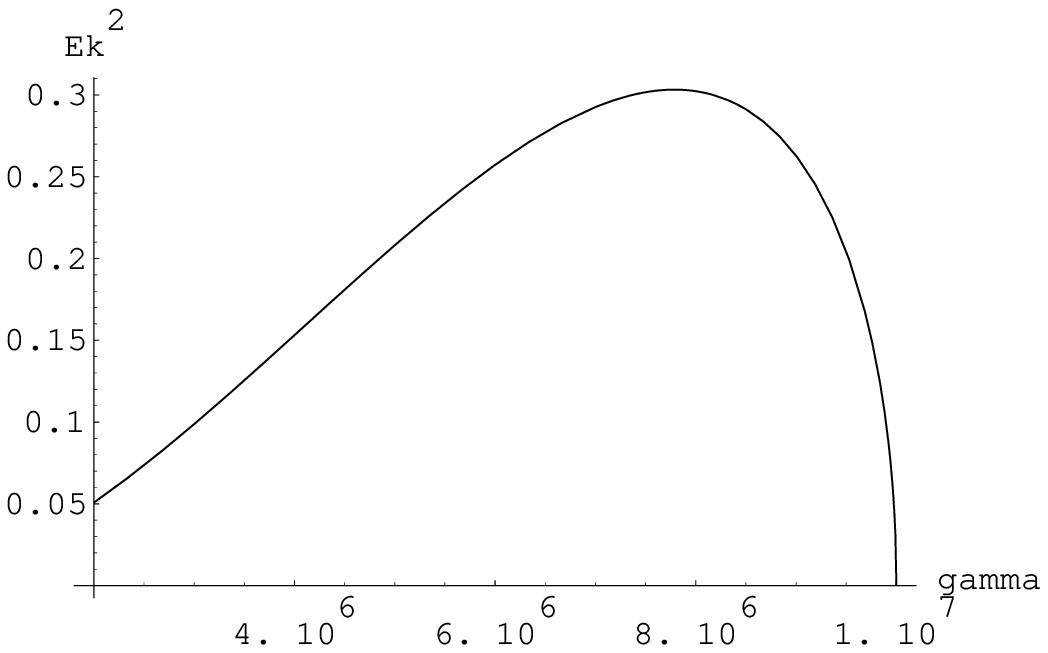,width=15.0cm}
\label{Ener11}
\psfig{file=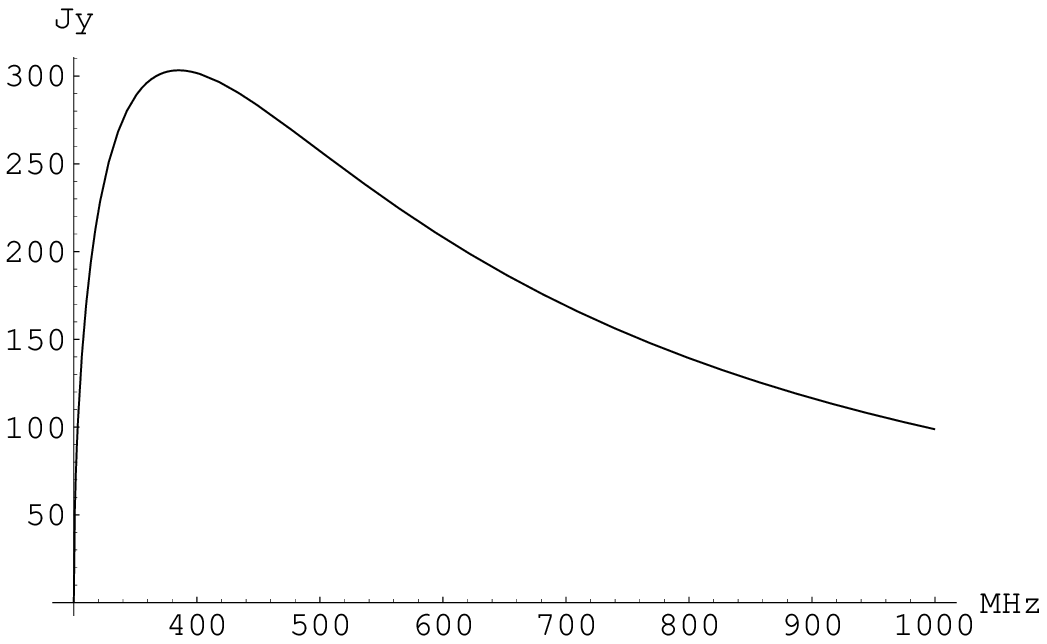,width=15.0cm}
\label{Ener22}
\caption{
Asymptotic one dimensional energy density in the waves in the 
$\gamma$-space (arbitrary units), and the predicted observed flux
in Janskys.
\label{Ener}}
\end{figure}

  It is noteworthy, that
a simple power law distribution for the spectral intensity and
distribution
function  cannot satisfy both Eqs. (\ref{222}) and (\ref{224}).
The particle  distributions function and the energy spectrum of the
waves are displayed in Figs. \ref{Distr} and \ref{Ener}.

We can now estimate the flux per unit frequency:
\begin{equation}
F(\nu)= 2 \pi  E_k^2 =
{ m c^4 \delta^3 \over \omega^2 r_e r_L r_S^2}
\left( {  \log({ \gamma_{max} \omega  r_L /( c \delta ) }) 
\over  \log( \gamma_{max}  / \gamma_{min} )  }  
\right)^{1/2},
\label{224b}
\end{equation}
characteristic pitch angle 
\begin{equation}
\psi_0 = \delta \left( { \pi  R_B r_L \over  r_S^2 }
\right)^{1/2} 
\left( { \log(\gamma_{max}/\gamma)  \over  \log (\gamma_{max}/\gamma_{min}) }
\right)^{1/4} \approx  10^{-6},
\label{226}
\end{equation}
(which remarkably stays  almost constant for a broad range of 
particles' energies and also for different values of $\gamma_{min}$),
and the 
total energy density in the waves
\begin{equation}
E_{tot} = \int_{\nu_{\min}}  ^{\nu _{\max}} F(\nu)  d \nu 
\approx 
{  m c^2 \gamma_{max} \over 4 \sqrt{\pi} r_e r_S^2 
\log^{1/2} (\gamma_{max}  / \gamma_{min} ) },
\label{224c}
\end{equation}

This total energy can be compared with the kinetic energy density
of the beam:
\begin{equation}
{ E_{tot} \over \gamma_b m c^2 n_{GJ} }
\approx \sqrt{ {\pi \over \log(\gamma_{max}  / \gamma_{min} ) }}
\label{224d}
\end{equation}
It means that some considerable fraction of the beam energy can be
transformed into waves.

We can also estimate the energy flux (\ref{224b}) at the Earth.
Assuming that distance to the pulsar is $ d\approx 1 $ kpc, we find 
\begin{equation}
F^{obs}(\nu) \approx  300 \, Jy  \left( {  \nu \over 400 {\rm MHz} }\right)^{-2}
\label{224e}
\end{equation}

With time, the value of $\gamma_{min}$ decreases as the particles are slowed
down by the radiation reactions force.
  Since at the given radius, the
particles with lower energies resonate with waves having larger frequencies,
more  energy will be transported to higher frequencies hardening
the spectrum. The lower frequency cutoff is determined by the
initial energy of the beam. No energy is transported to  frequencies
lower than 
\begin{equation}
\omega_{min} = {\omega_B \over \gamma_b \delta}
\label{227}
\end{equation}
This simple picture, of course, will be modified due to the 
propagation of the flow in the inhomogeneous magnetic field of pulsar
magnetosphere.

\section{Conclusion}
\label{Conclu}

In this work we investigated the new saturation mechanism
for the cyclotron-Cherenkov instability of a beam in a 
inhomogeneous magnetic field. We showed that for the typical parameters 
of the pulsar magnetosphere it is possible to reach quasisteady 
state, in which the transverse motion of particles is determined by the
balance of a radiation reaction  force due to the emission 
at anomalous Doppler effect and the force arising in the inhomogeneous
magnetic field due to the conservation of adiabatic invariant.
The resulting wave intensities are sufficient to explain the 
observed fluxes from radio pulsars.

\acknowledgements

I would like to thank Roger Blandford, Peter Goldreich  and Gia Machabeli
for their useful 
comments.

\begin{table}
\caption{Dimensions of the main used quantities}
\begin{tabular}{|l|c|c|c|c|c|c|c|c|c|c|}
\hline
$\phantom{ {{{ {a\over b} \over {a\over b} } } \over {{ {a\over b} \over {a
\over b} } } } }$
$E ^2({\bf k})$ &
 $ E^2_k$ & 
$   n({\bf k}) $ &
 $ n(k)  $ 
 & $ \alpha$, $\beta_R$  &
 $ D_{\psi \psi} $&   
$D_{\psi p}$ & 
$D_{p p }$ &
 $ D$ & 
$A$&
 $ f(p) p^2 dp,\, f(\gamma) d \gamma $ 
 \\ \hline
$\phantom{ {{{ {a\over b} \over {a\over b} } } \over {{ {a\over b} \over {a
\over b} } } } }$
erg  &
$ {\rm {erg \over cm^2}} $ &
 1  &
$ { 1 \over cm^2} $ &
${\rm {erg \over cm }} $ &
  $ {\rm { 1 \over sec }} $ &
$ {\rm { erg  \over cm  }} $ & 
$ {\rm  { erg^2 \,sec \over cm^2 }} $ &
 ${\rm { cm^2 \over erg  \, sec } }$ &
$ {\rm {  cm }}$ &
1\\ 
\hline
\end{tabular}
\label{Dimen1}
\end{table}

\end{document}